\documentclass{appolb}
\usepackage{graphicx}

\begin{document}
\title{MONTE CARLO SIMULATION OF ACTIVE NEUTRON INTERROGATION SYSTEM DEVELOPED FOR DETECTION OF ILLICIT MATERIALS%
\thanks{Presented at 2nd Jagiellonian Symposium on Fundamental and Applied Subatomic Physics, Krakow, Poland}%
}
\author{ A.Sh. Georgadze
\address{Kiev Institute for Nuclear Research\\
	Kiev 03680, Ukraine\\
	E-mail: a.sh.georgadze@gmail.com, georgadze@kinr.kiev.ua }
}
\maketitle
\begin{abstract}
Nowadays, the current threat of international terrorism is set to a severe level, demanding worldwide enhanced security. Radioactive materials that could be fashioned into a radiation dispersal device typically emit gamma rays, while fissile materials such as uranium and plutonium emit both neutrons and gamma rays via spontaneous or induced fission. Therefore, the detection and identification of hazardous materials has become increasingly important. We present the results of GEANT4 Monte Carlo simulation of an active neutron interrogation system based on highly segmented neutron/gamma-ray detector and pulsed neutron generator. This system is capable of detecting and imaging radioactive and special nuclear materials, explosives and drugs. The segmented scintillation detector works as a scatter camera, allowing selection of a gamma ray events that undergo multiple interaction in detector blocks for radioactive source localization. The detector consist of blocks made of plastic scintillator which provide scattering and blocks of CsI, used as an absorber, used as an absorber, which has to be efficient to detect the characteristic gamma radiation for the identification. Because of this imaging capability background events can be significantly rejected, decreasing the number of events required for high confidence detection and thereby greatly improving its sensitivity. A scatter imager for the detection of shielded radioactive materials has been conceptualized, simulated, and refined to maximize sensitivity while minimizing cost.
\end{abstract}
DOI:10.5506/APhysPolB.48.1683
  
\section{Introduction}
Lately the threat of international terrorism grows constantly, requiring the increased safety in a whole world. Presently disorganization of state administration, task of economic and political harm, compulsion of power are the primary purposes of international terrorism to the change of policy. Bomb terrorism is one of the most widespread. Therefore detection of explosives and threat materials inside luggage of an aircraft passengers has become an essential requirement.

For the search of the hidden treads systems based on the direct methods on the use of X-rays, vapor detection and sniffer dogs or on the use of nuclear quadrupole resonance have proved their efficiency \cite{berm1, jank1, singh1}. However, hermetical casing or impenetrable metallic shell make impossible exposure of the hidden hazardous substances and limit the possibility of their detection. In such cases the use of radiation with high ability to penetrate and detect characteristic signs of interaction with chemical elements of hazardous substances is necessary.

Prompt gamma-ray neutron analysis \cite{pese, Buff} has been investigated for over six decades as possible solution offering an on-stream, non-destructive, rapid method for the luggage inspection at the airports. Objective of this type of inspection is to determine small quantity of explosive of the order 200 g. The basic problem faceing by this technique is that, the principal elements which constitute explosives and illicit drugs (H, C, N and O) are present also in the common materials, but with different concentrations. High density and somewhat increased concentration of nitrogen is a characterization of explosive substances. In addition, the background associated with gammas, produced by neutron interaction with surrounding materials mask the signals form explosives. So, the spatial methods have to be applied to reduce backgrounds to an acceptable level. 

To overcome background different methods were developed. One approach provides Coded aperture imaging apparatus and methods for the detection and imaging of radiation which results from nuclear interrogation of a target object. This method allow to localize object and decrease background but on the expense of gamma ray signal intensity loss. Additional shortcoming of this method is that changing the coded aperture masks is required, that resulted in time duration increase for an inspection. Another approach is the Associated Particle Imaging Method \cite{hurl1}. In this case the characteristic gamma radiation is recorded in coincidence with the signal from detector which register a particles from neutron production reaction $d~+~t~\rightarrow~\alpha~+~n$. This method allow tomography of the explosive object. The only shortcoming of this method is a high prices of the neutron generator with $\alpha$-particle detector embedded in. 

In this study, we have performed Monte Carlo simulation with the help of GEANT4 package \cite{agos} neutron interrogation with fast 14.1 MeV neutrons intended to check the luggage of aircraft passenger on presence of explosive materials. For the detailed test on presence of explosives in the luggage an electronic collimation method based on Compton scattering tracking is developed for the reduction of background from gamma rays, produced by neutron interactions with shielding materials.

\section{Design of detection system}
Detector (Fig. 1) consist of plastic scintillation rectangular bars with size $5~\times~5~\times~20$~cm$^3$ covered by reflecting film of aluminum-backed Mylar and film that contains gadolinium oxide and bars made of CsI or BGO inorganic scintillator. Layer of air between scintillation bars and Mylar ensures full internal scattering of photons. Bars should be connected into one unit of $15~\times~50~\times~20$~cm$^3$ dimensions consisting of two rows of plastic scintillator bars serving as a scatter, and last raw composed from inorganic scintillator for high- efficiency absorption. Additionally plastic scintillation bars are covered by film containing gadolinium oxide paint to allow neutron detection. Such a design will  make possible to select gammas which have Compton scattered in plastic and absorbed in CsI scintillator. Since gamma-rays with energy more then 3 MeV are scattered on small angle the backprojection method allow to reconstruct the source location. 

\begin{figure}[htb]
	\centerline{%
		\includegraphics[width=6.5cm]{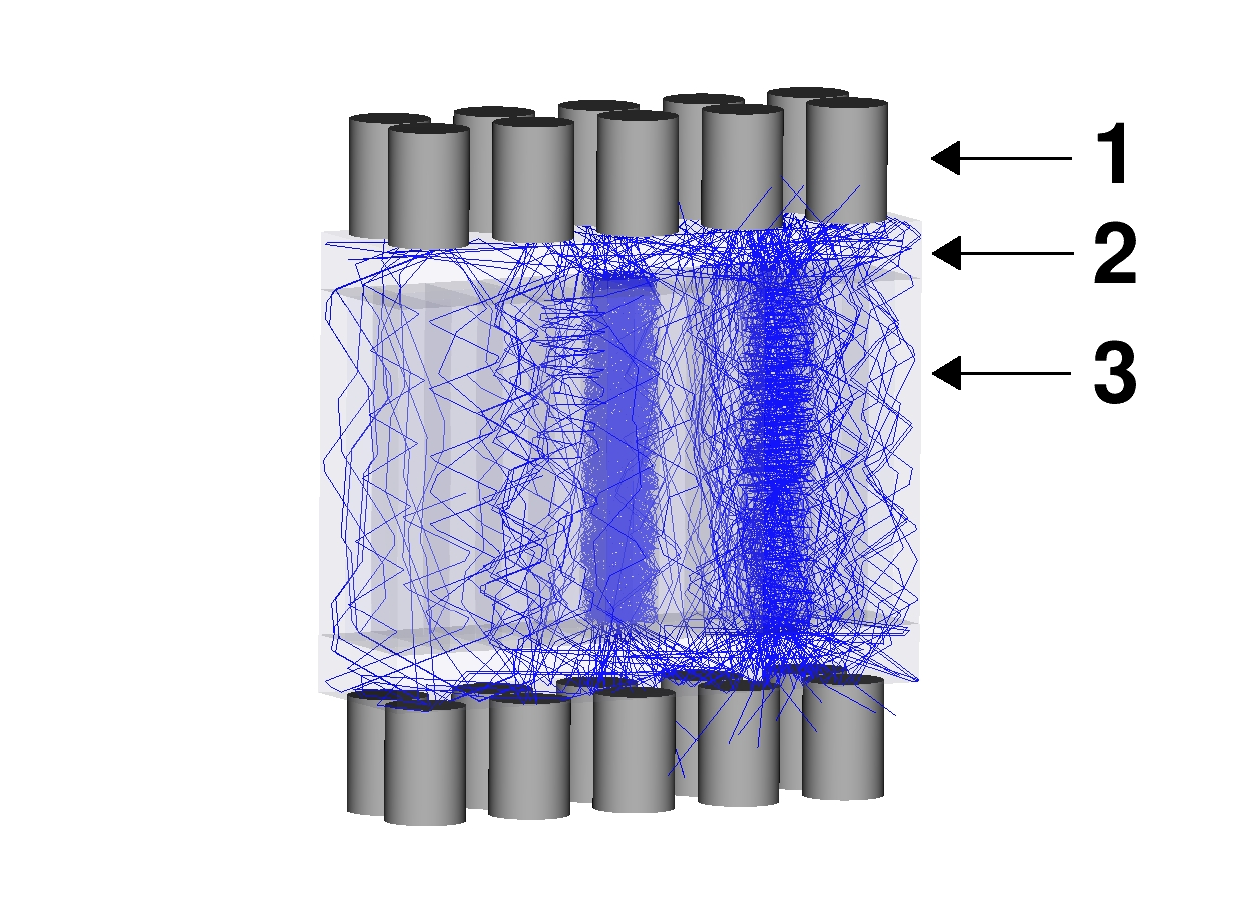}}
	\caption{Gamma-neutron detector scheme (rendered picture): 1 -- PMT; 2 -- light sharing guide; 3 -- scintillation bars.}
	\label{Fig:1}
\end{figure}

\begin{figure}[htb]
	\centerline{%
		\includegraphics[width=6.5cm]{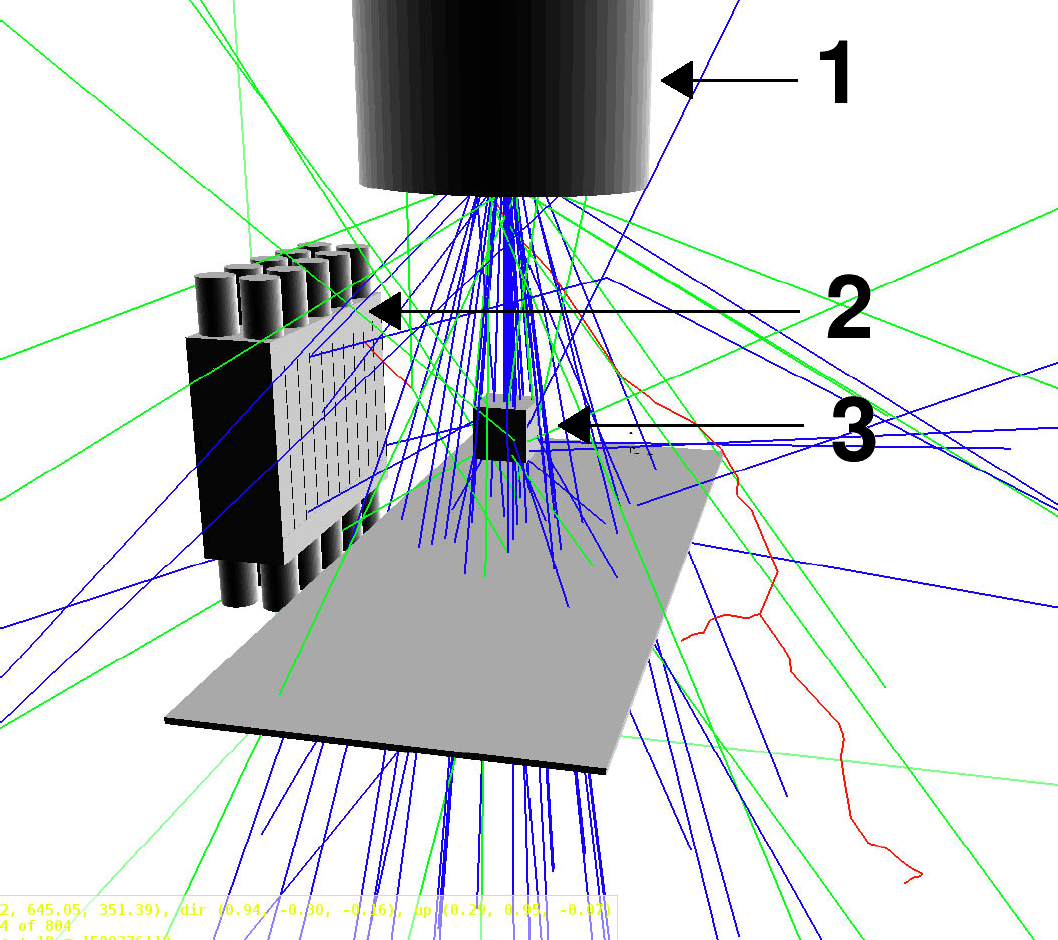}}
	\caption{The 3-D visualization of GEANT4 simulation of of neutron interrogation setup and particle tracking by VRML: 1-neutron generator in shield housing from lead and borated polyethylene, 2 -- detector, 3 -- inspected object.}
	\label{Fig:2}
\end{figure}

Scintillation bars are sandwiched on both sides by continuous light guides 5 cm thick. At each side 10 photomultiplier tubes (PMT) are placed on top of light guides to collect scintillation light. The light sharing guides allow scintillation light produced in one or two bars to be transferred and shared between all 20 PMTs almost simultaneously, forming scintillation photon distribution pattern $\{\textit{m}_1\textit{;m}_2\textit{; : : : ;m}_{20}\}$. The detector design proposed is based on application of Anger-logic algorithm which is widely applied in detectors used for Positron Emission Tomography. 
The trajectories of the particles under simulation, as well as the simulation geometry, can also be visualized using the VRML visualization driver (see Fig. 2 for illustration).

\section{Detector geometry simulation}
Full simulations of the detector were performed using GEANT4 package \cite{agos}. The scintillation bars supposed to be made of PS-923A \cite{arti} plastic scintillator produced by Amcrys company (Ukraine). They were modeled as 95\% polished and 5\% diffuse (Lambertian). The optical properties of aluminized Mylar film were taken from the RealSurface1.0 data set \cite{jane} of Geant4 (PolishedVM2000). The data on bulk attenuation length (BAL) of the scintillation 250-450 cm and light output 59\% of anthracene, that correspond to $10^4$ optical photons/MeV were taken from [8]. The PMTs were modeled as a 1 mm thick disk with 100\% active area, made of borosilicate glass BK7 with refractive index n~=~1.52. The light detector was placed on outer surface of the PMT glass BK7 and assigned as ideal absorber with 100\% active surface and 25\% efficiency to detect optical photons.

\section{Simulation of explosive detection}

The Monte Carlo Geant4 Transport code has been used for the present simulations. A D-T pulsed neutron generator operating at $10^8$ neutrons per second has been simulated as fixed point isotropic neutron source. The neutron source was located inside the shield made of lead and borated polyethylene with opening in form of cone, which serves as neutron flux collimator to ensure that all area of 50~$\times$~50 cm$^2$ in size where luggage is located will be covered. The emitted 14.1 MeV fast neutrons produce high energy gamma ray signals as a result of the inelastic interaction of fast neutrons with carbon, oxygen and nitrogen nuclei present in an explosive material. The radiation shield made of lead and concrete is surrounding the detector and volume where the luggage inspection take place.

Simulations were implemented in several stages. In the first stage the Iron was considered as a shielding material. But in this case large continuous background do not allow to resolve gamma lines. Then lead was used as a shield. After careful choosing of shielding geometry the gamma lines from C, N and O have became possible to resolve.

On Fig. 3 and Fig. 4 we presented gamma spectra of simulation of two second exposure to 14.1 MeV neutron flux of explosive RDX (C$_3$H$_6$N$_6$O$_5$), and an explosive simulant melamin (C$_3$H$_6$N$_6$) of 5 $cm^3$ in volume. To take into account impact of common harmless materials on explosive detection we have simulated the explosive placed inside the bag of $20~\times~20~\times~30$ cm$^3$ in size, filled with clothes simulated as a cotton of 0.2 g/cm$^3$ density, composed from cellulose cellophane and air, but no considerable impact on detected spectra is found. 

As seen from Fig. 3, the energy distribution is characterized by three peaks in the spectrum: a peak in the total gamma energy absorption from the line $E_{\gamma}=4,43$~MeV from $^{12}$C, a peak with $E_{\gamma}=5.10$~MeV from $^{14}$N and peak with $E_\gamma=6.13$~MeV form $^{16}$O. The RDX spectrum (in blue color) has characteristic gamma with $E_\gamma= 6.13$~MeV form $^{16}$O, but there is no such a line present in the gamma spectrum of explosive simulant -- melamin (in red color). Another explosive materials used for the simulation are: ammonium nitrate (H$_4$N$_2$O$_3$), TNT (C$_7$H$_5$N$_3$O$_6$) and C4 (C$_4$H$_6$N$_6$O$_6$).

\begin{figure}[htb]
\centerline{%
\includegraphics[width=6.5cm]{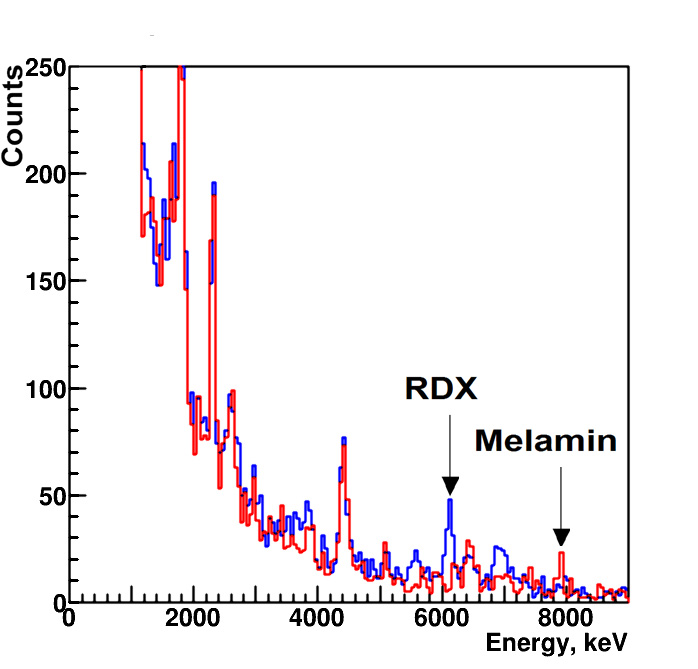}}
\caption{The energy spectra of  the RDX (blue) and Melamin (red)}
\label{Fig:3}
\end{figure}

\begin{figure}[htb]
	\centerline{%
		\includegraphics[width=6.5cm]{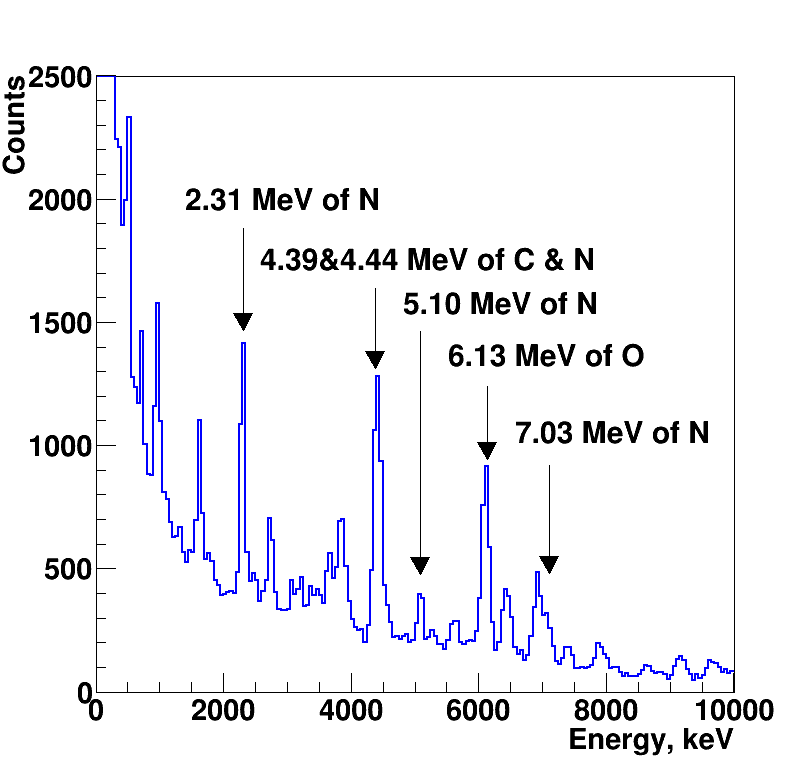}}
	\caption{The energy spectra of  the RDX after 120 seconds of testing with spatial selection of gammas.}
	\label{Fig:4}
\end{figure}

\section{Electronic collimation}

In short inspections during several second when the luggage is moving through the detection area the precise chemical composition can not be retrieved.  In case of suspicion that luggage is containing illicit material the luggage can be stopped at the inspection area and additional testing can be performed during several minutes with applied electronic collimation for background suppression.

\begin{figure}[htb] \label{ fig5} 
	\begin{minipage}[b]{0.5\linewidth}
		\includegraphics[width=.95\linewidth]{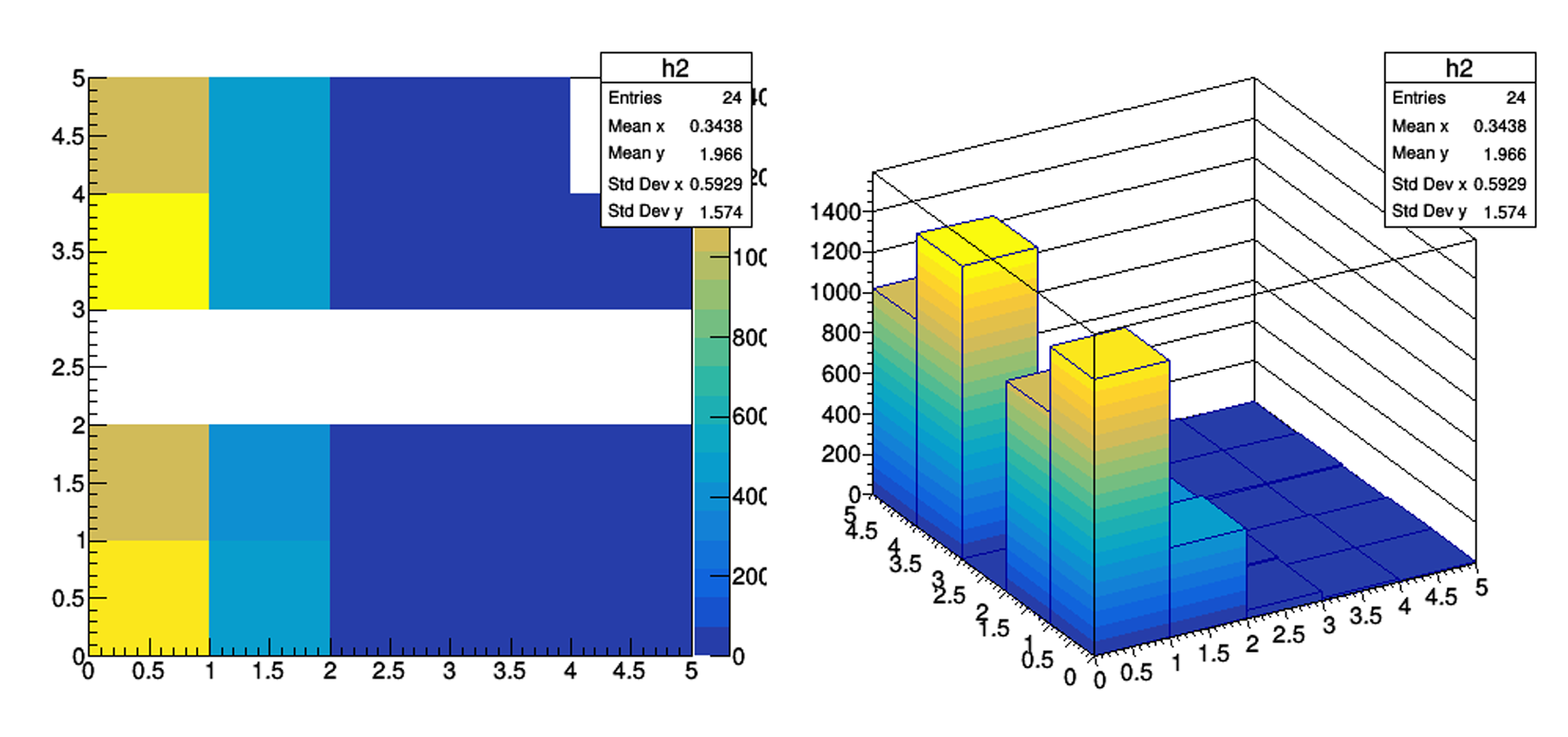} 
	\end{minipage} 
	\begin{minipage}[b]{0.5\linewidth}
		\includegraphics[width=.95\linewidth]{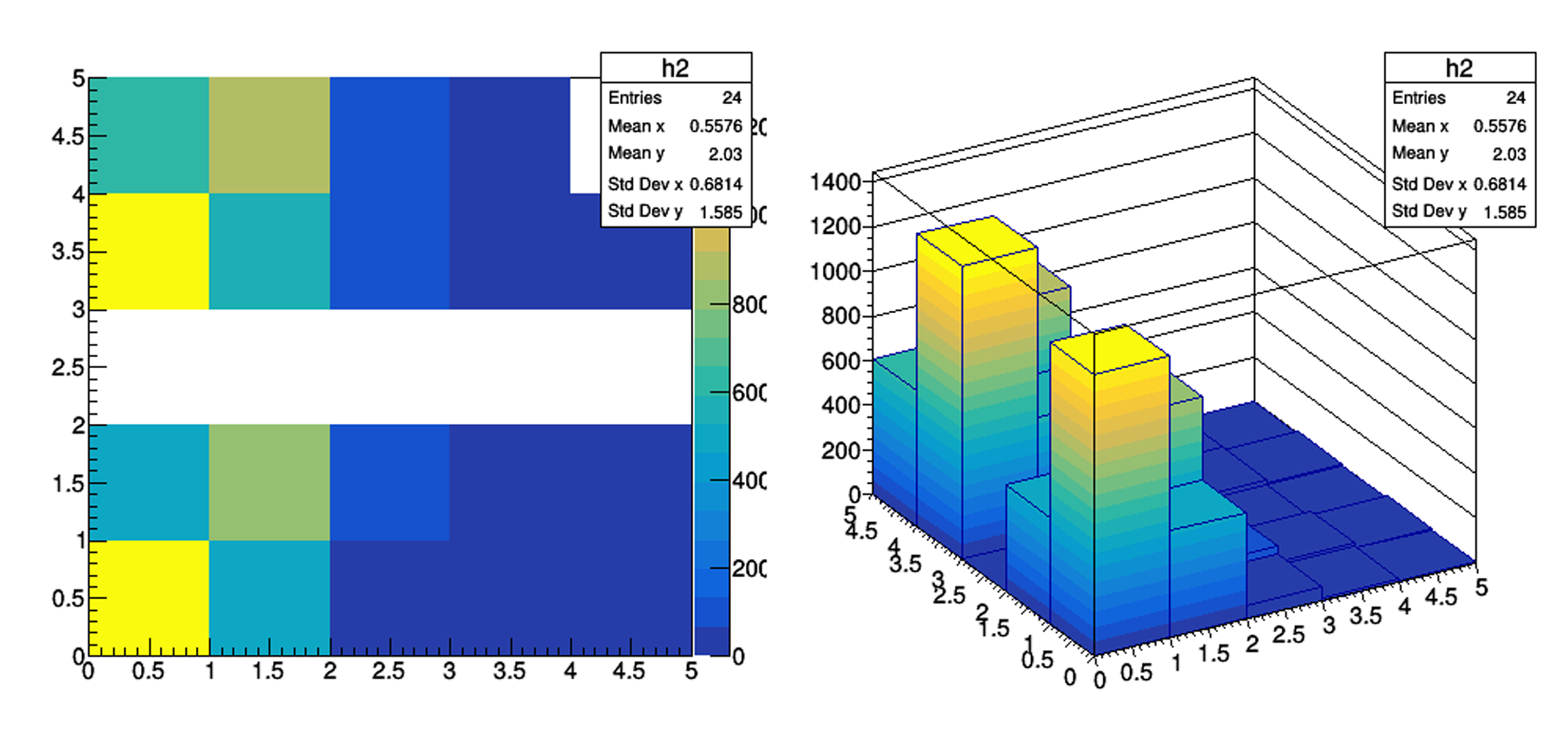} 
	\end{minipage} 
	\begin{minipage}[b]{0.5\linewidth}
		\includegraphics[width=.95\linewidth]{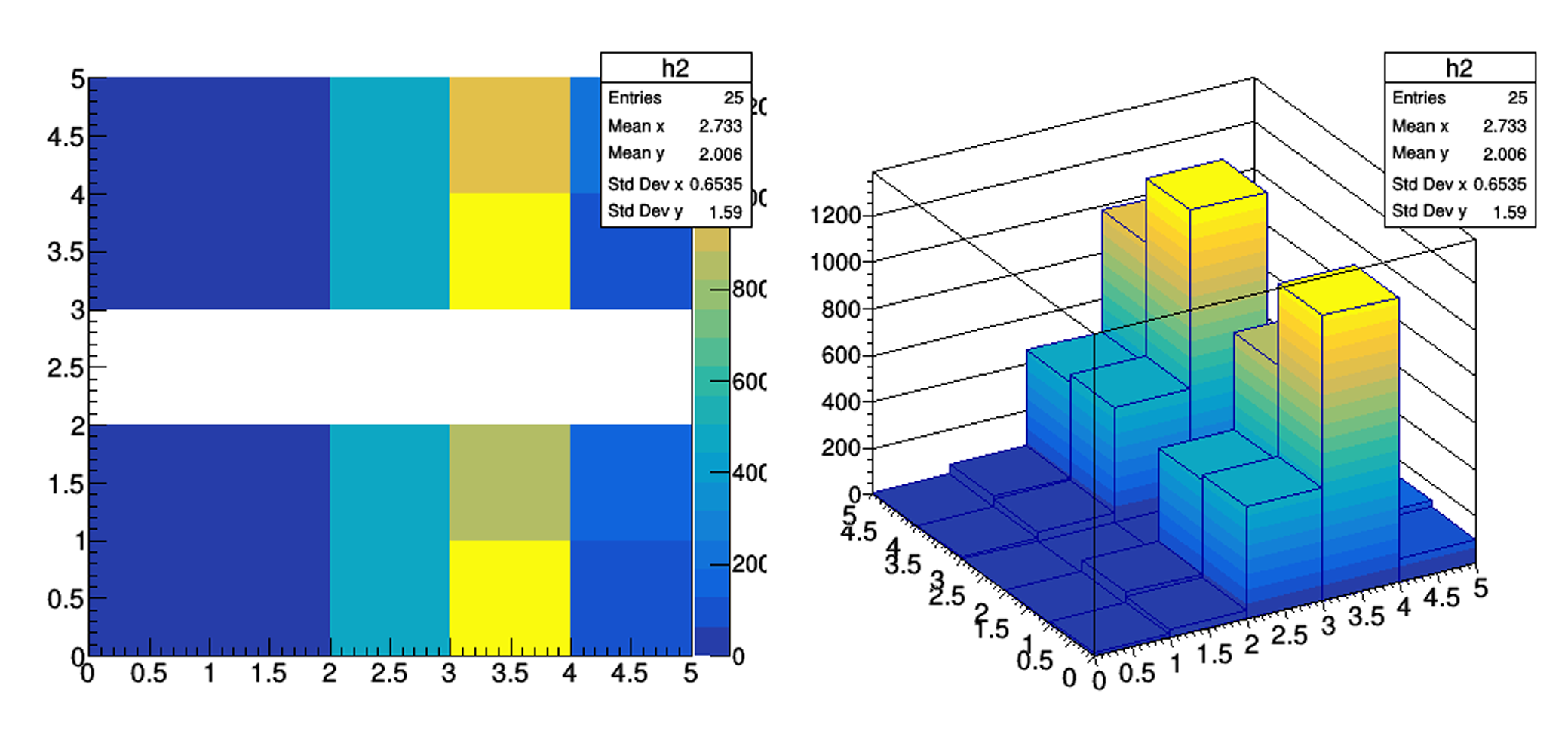} 
	\end{minipage}
	\hfill
	\begin{minipage}[b]{0.5\linewidth}
		\includegraphics[width=.95\linewidth]{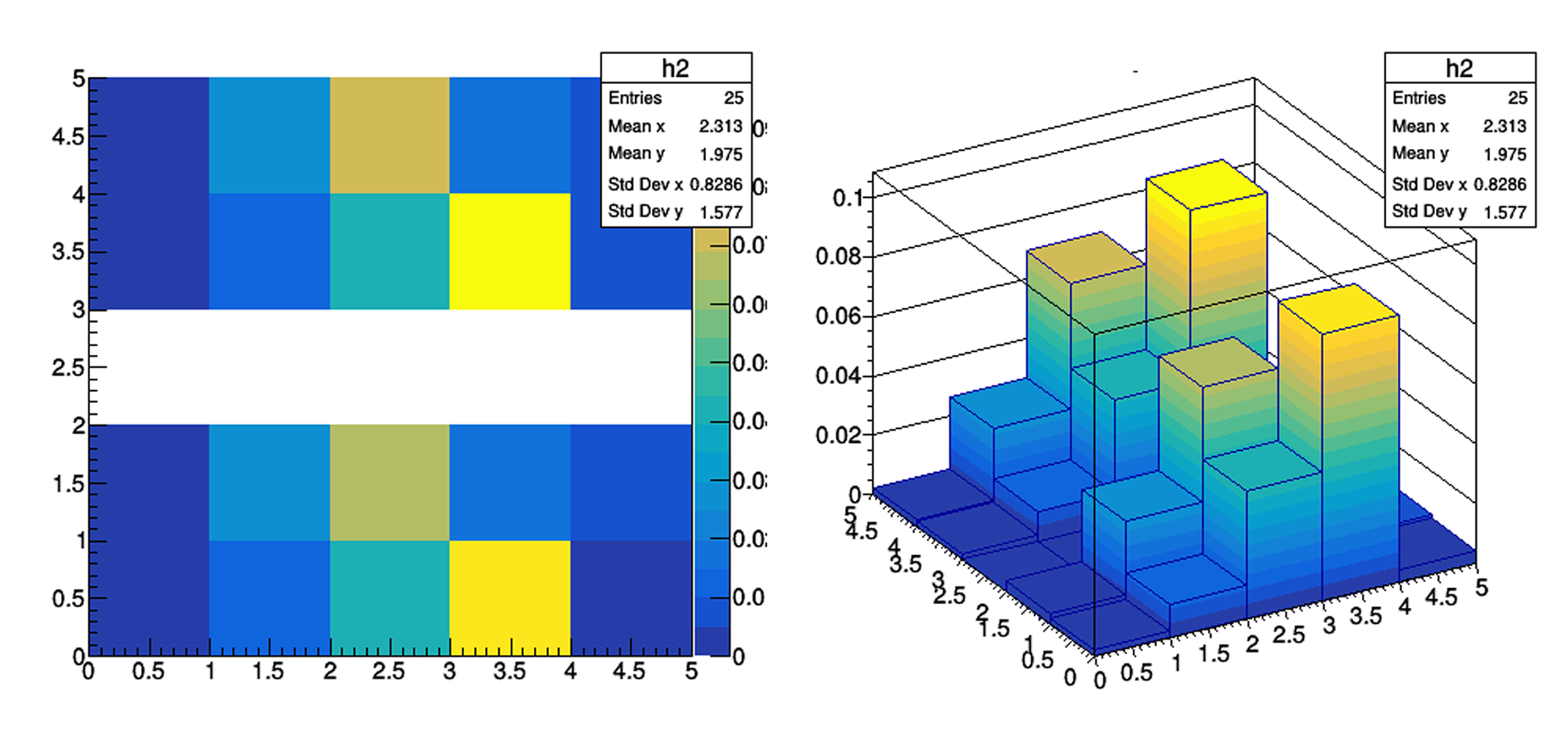} 
	\end{minipage} 
			\caption{Photoelectron distribution patterns $\{\textit{m}_1\textit{;m}_2\textit{; : : : ;m}_{20}\}$. The left column indicates gamma ray orthogonal to detector surface and right column indicates gamma ray beam hitting the detector surface under the  angle of 10$^o$.}
			\label{Fig:5}
\end{figure}

The simulation of gamma ray transport in the detection system results in conclusion that approximately 16\% of gammas produced by neutrons interactions with inspected object are scattered first in plastic scintillation bar and absorbed in CsI scintillator allowing reconstruction of source location. This is due to the fact that high energy gammas of several MeV's are scattered mainly on small angles. Due to segmented design the effect of Compton scattering can be applied to select gammas coming from only inspected area. For the source localization only coincidence events are selected. Coincidence events are defined as events with non-zero energy deposit in exactly one cell in either one of the scatter plastic scintillation bar, and exactly one in the absorber layer. This ensures the events have gone through a Compton scatter but also minimizes the chance of multiple-scattering events.

The effect of electronic collimation is illustrated on Fig.5. Gamma quanta coming from different directions resulted in considerably different scintillation photon distribution patterns. To be able to reconstruct the  position of the gamma source in the inspection volume, the detector response have to be calibrated as a function of gamma source position. This 3D simulated calibration set was then used to select gammas coming from inside the inspected area by Maximum Likelihood Estimation (MLE). 

The energy spectra of  the RDX after 120 seconds of additional testing with spatial selection of gammas is presented on Fig.4. One can see that all gamma peaks from C, N and O are resolved and their intensities can be used for chemical analysis of inspected substances. 

\section{Method of delayed coincidences} 

The detection of concealed Special Nuclear Material (SNM) in the airport passengers luggage is an important goal for prevention of a nuclear terrorism. Fast neutrons emitted in the neutron induced fission on SNM can be detected with plastic scintillator combined with Gd lined neutron-gamma converter. Neutron detection and neutron-gamma discrimination can be done with delayed-coincidence signature of two time correlated pulses, the so called capture-gated method \cite{Knoll}. The prompt signal is produced by fast neutron scattering on protons in hydrogen reach plastic scintillator material and delayed signal is produced by capture of thermalized incident neutron on Gadolinium nucleus. The latter reaction leads to a cascade of 2-3 gamma-rays emitted with total energies of 8.46 MeV ($^{155}$Gd) and 7.87~MeV ($^{157}$Gd) that are essentially bigger than energy of events from natural radioactive background with upper limit of 3 MeV. Gd in natural abundance can be used as active n-$\gamma$ converter due to high abundance of isotopes $^{157}$Gd (15.7\%) and $^{155}$Gd (14.7\%). 

Another benefit of using this isotopes is the highest thermal neutron capture cross-sections among all known nuclides, $\sigma_{cap}$ = 254000 b and $\sigma_{cap}$ = 61000 b, respectively. In simulations we have considered neutron-gamma converter as a thin film placed in between scintillation bars and made of acrylic (PMMA) mixed with gadolinium oxide (Gd$_2$O$_3$) with a Gd/scintillator weight ratio of 0.5$\%$.

Scintillation bars geometry impact considerably on neutron detection efficiency. With larger cross section of scintillation bar there is the high probability of neutron to be captured by Hydrogen with emitting 2.2 MeV gamma quantum which is difficult to detect due to high counting rate in this energy region from natural background detection. Smaller cross section results in increased light absorption and energy resolution degradation.  For chosen geometric dimensions of scintillation bars simulation results in probability of neutron capture on gadolinium of 0.88 and of an average time of neutron capture - 26 $\mu$s. The efficiency of neutron detection is estimated to be about 10-15\%. 

\section{Enriched uranium detection}

Modeling of detector function response to neutrons and gammas with Monte-Carlo method shows that high effectiveness for gammas and neutrons detection at the same time can be reached and discrimination between these two particle types can be performed by application of capture-gated method based on delayed coincidence of neutron scattering at protons of plastic scintillator and reaction of neutrons capture by $^{155}$Gd and $^{157}$Gd nuclei.

Discrimination between gammas and neutrons can be reached due to different quantities of scintillation bars that detected events dealing with interaction of secondary particles with detector material. Preliminary modeling of detector response with application of software code GEANT4 showed that detection effectiveness of neutrons from SNM will be about 10\%. 
Although most fissile materials naturally emit neutrons and/or $\gamma $-rays, the intensity of the spontaneous radiation is low, and the energies of the $\gamma $ -rays are fairly low in most cases. The active neutron interrogation approach involves bombarding the sample with external neutron source and thereby inducing additional fission in the sample and counting the emitted neutrons due to fission. 

To test sensitivity of the system to presence of fissile materials the enriched uranium sample consisting of 10\% of $^{238}$U and 90\% of $^{235}$U with the mass of 200 g was placed in the center of system inside the bag filled with clothes. The clothes were simulated as 0.2 g/cm$^3$ Cellulose cellophane material taken from Geant4 material library mixed with air.  The sample inside the luggage was irradiated with $10^8$ of 14.1 MeV neutrons, placed inside shield-collimator. Presently Differential Die-Away (DDA) active neutron interrogation technique was not used for fissile content determination. Instead of that the considerable growing of integral counting rate in the whole energy range with characteristic peaks of full absorption of gammas from neutron capture on gadolinium is an indication of presence of enriched uranium. 

\section{Conclusions}

According to preliminary modeling results of detector parameters, it was determined that there is possibility to improve explosive detection effectiveness at the account of electronic collimation for source localization in space resulting in essential background reduction. For the neutron generator intensity of  $10^8$~n/s, the 400~g of explosives hidden in a luggage can produce alarm on presence of suspicious material. In a few minutes of the additional test with applying the electronic collimation the chemical composition of the object can be established. Additionally, 200 g of enriched uranium sample can be found with the help of active neutron interrogation.

\end{document}